\documentclass[a4paper,12pt]{article}

\usepackage{amsmath}
\usepackage{hyperref}
\usepackage{graphicx}
\usepackage{cleveref}
\usepackage{booktabs}
\usepackage{esint}
\usepackage{amsfonts}
\usepackage{xspace}
\usepackage{fontawesome}

\usepackage{geometry}
\usepackage[skip=12pt]{caption}

\usepackage[T1]{fontenc}
\usepackage[utf8]{inputenc}
\usepackage{biolinum}

\usepackage{fourier}
\usepackage[scaled=0.83]{beramono}
\usepackage{microtype}

\usepackage{titlesec}
\titleformat{\section}{\sffamily\Large\bfseries}{\sffamily\Large\bfseries\thesection}{0.5em}{\sffamily\Large\bfseries}
\titleformat{\subsection}{\sffamily\large\bfseries}{\sffamily\large\bfseries\thesubsection}{0.5em}{\sffamily\large\bfseries}

\usepackage[numbers,sort&compress]{natbib}
\usepackage[english]{babel}
\makeatletter
\def\NAT@spacechar{\,}
\makeatother

\hypersetup{
    colorlinks=true,
    allcolors=[rgb]{0.26,0.41,0.88},
}
   
\newcommand{\oleft}{\mathopen{}\mathclose\bgroup\left}
\newcommand{\oright}{\aftergroup\egroup\right}

\newcommand{\hvp}{\textsc{hvp}}
\newcommand{\damuhad}{\ensuremath{a_\mu^\hvp}}
\newcommand{\dalphahad}{\ensuremath{\Delta \alpha_\text{had}}}
\newcommand{\linkemail}[1]{\href{mailto:#1}{#1}}
\newcommand{\code}{\textsf}
\newcommand{\refcite}{Ref.~\cite}
\newcommand{\gev}{\,\text{GeV}}
\newcommand{\var}[1]{\code{Var}\oleft[\vphantom{R_i, R_j}#1\oright]}
\newcommand{\cov}[1]{\code{Cov}\oleft[\vphantom{R_i, R_j}#1\oright]}
\newcommand{\mean}[1]{\code{E}\oleft[\vphantom{R_i, R_j}#1\oright]}
\newcommand{\corr}[1]{\rho\oleft[\vphantom{R_i, R_j}#1\oright]}
\newcommand{\benchmarki}{\code{KNT18}~\cite{Keshavarzi:2018mgv}\xspace}
\newcommand{\benchmarkii}{\code{KNT19}~\cite{Keshavarzi:2019abf}\xspace}
\DeclareMathOperator*{\argmax}{arg\,max}
\newcommand{\gh}[1]{\href{https://github.com/#1}{\faGithub}}

\begin{document}

\thispagestyle{empty}
\renewcommand{\thefootnote}{\fnsymbol{footnote}}

\begin{center} 
{\huge\bf\sffamily\boldmath Modeling the $R$-ratio and hadronic contributions to $g-2$ with a Treed Gaussian Process}
\end{center}
\vspace{1mm}
\begin{quote}
\begin{center}
{\begin{large}\textsf{\textbf{Andrew Fowlie}}\thanks{\textsf{\linkemail{andrew.fowlie@xjtlu.edu.cn}}}\end{large}\\[2mm]
{\textit{Department of Physics, School of Mathematics and Physics, Xi'an Jiaotong-Liverpool University,  Suzhou 215123, China}\\[3mm]}
\begin{large}\textsf{\textbf{Qiao Li}}\thanks{\textsf{\linkemail{qiaoli@njnu.edu.cn}}}\end{large}}\\[2mm]
{\textit{Department of Physics and Institute of Theoretical Physics, Nanjing\\ Normal University, Nanjing, Jiangsu 210023, China}\\}
\end{center}
\end{quote}
\vspace{1mm}
\begin{quote}
The BNL and FNAL measurements of the anomalous magnetic moment of the muon disagree with the Standard Model (SM) prediction by more than $4\sigma$. 
The hadronic vacuum polarization (HVP) contributions are the dominant source of uncertainty in the SM prediction.
There are, however, tensions between different estimates of the HVP contributions, including data-driven estimates based on measurements of the $R$-ratio.
To investigate that tension, we modeled the unknown $R$-ratio as a function of CM energy with a treed Gaussian process (TGP). 
This is a principled and general method grounded in data-science that allows complete uncertainty quantification and automatically balances over- and under-fitting to noisy data.
Our tool yields exploratory results are similar to previous ones and we find no indication that the $R$-ratio was previously mismodeled.
Whilst we advance some aspects of modeling the $R$-ratio and develop new tools for doing so, a competitive estimate of the HVP contributions requires domain-specific expertise and a carefully curated database of measurements.
\gh{qiao688/TGP_for_g-2}
\end{quote}

\makeatletter
\@thanks
\makeatletter
\newpage
\renewcommand{\thefootnote}{\arabic{footnote}}
\setcounter{page}{1}
\setcounter{footnote}{0}

\section{Introduction}

The final measurement of the anomalous magnetic moment of the muon at the Brook\-{}haven National Laboratory (BNL) E821 experiment~\cite{Muong-2:2006rrc}
differed from the Standard Model (SM) prediction by $3.7\sigma$. This discrepancy was replicated by the Fermi National Accelerator Laboratory (FNAL) E989 measurement~\cite{Muong-2:2021ojo}. The combined measurement from BNL and FNAL,
\begin{equation}
a_{\mu } = 116\,592\,061\,(41) \times 10^{-11}, 
\end{equation}
deviates from the SM theory prediction by $4.2\sigma$, motivating the possibility of physics beyond the SM~\cite{Athron:2021iuf} as well as scrutiny of the SM prediction~\cite{Aoyama:2020ynm}. 

The SM prediction for $a_{\mu}$ incorporates contributions from quantum electrodynamics (QED), electroweak interactions (EW), and hadronic effects~\cite{Athron:2022qpo}. While QED and EW contributions can be calculated with high precision using perturbation theory and are well-controlled~\cite{Teubner:2010ah}, hadronic contributions are harder to compute and are the largest source of uncertainty in predictions for $a_{\mu}$. The hadronic contribution can itself be decomposed into the following parts:
\begin{align}
a_{\mu }^{\mathrm{had}}&=\damuhad+a_{\mu }^{\mathrm{LbL}}, 
\end{align}
where $\damuhad$ is the hadronic vacuum polarization (HVP) contribution and $a_{\mu }^{\mathrm{LbL}}$
is the light-by-light scattering contribution.
In this work we focus on the leading-order (LO) contribution to HVP. This is particularly challenging and the dominant source of uncertainty, making up about $80\%$ of the total. The two most popular computational methods are lattice QCD and estimates from dispersion integrals and cross-section data. 

Lattice methods calculate the HVP contribution by discretizing spacetime and performing a weighted integral of relevant functions over Euclidean time. This requires significant computational resources and currently cannot match the nominal precision achieved by data-driven methods. Data-driven methods use data from, for example, the KLOE~\cite{KLOE-2:2017fda}, BaBar~\cite{BaBar:2018erh,BaBar:2007qju,BaBar:2017zmc,BaBar:2018rkc}, SND~\cite{SND:2016drm} and CMD-3~\cite{CMD-3:2017tgb,CMD-3:2019ufp} experiments to estimate the $R$-ratio,
\begin{equation}
R(s) \equiv \frac{\sigma^{0}(e^+e^- \to \text{hadrons})}{\sigma(e^+e^- \to \mu^+\mu^-)}.
\end{equation}
This is a function of the center-of-mass (CM) energy, $\sqrt s$.\footnote{The superscript in ${\sigma^{0}}$ denotes the bare cross section for $e^{+}e^{-}$ annihilation to hadrons, which is defined as the measured cross section that has been corrected for electron-vertex loop contributions, initial state radiation (ISR) and vacuum polarization (VP) effects in the photon propagator.} 
From the estimate of the $R$-ratio, the leading-order (LO) HVP contributions can be computed through the dispersion integral,
\begin{equation}
\damuhad = \frac{\alpha^{2}}{3\pi^{2}} \int\limits_{m_\pi^2}^\infty \frac{K(s) R(s)}{s} \, \text{d}s
\label{eq:HVP}
\end{equation}
where $K(s)$ is the QED kernel~\cite{Lautrup:1968tdb,Brodsky:1967sr}. Hadronic contributions to the effective electromagnetic coupling constant at the $Z$ boson mass can be computed in a similar way through the dispersion relationship
\begin{equation}
\dalphahad = \frac{\alpha M_Z^{2}}{3\pi}\fint\limits_{m_{\pi}^{2}}^{\infty}\frac{R(s)}{s(M_Z^{2}-s)}\text{d}s
\label{eq:alpha_had}
\end{equation}
where $\fint$ represents the principal-value prescription. There is growing tension between results from the two approaches~\cite{Wittig:2023pcl,benton2023datadriven}. A recent lattice QCD calculation found~\cite{Borsanyi:2020mff}
\begin{equation}
    \damuhad = 707.5\, (5.5) \times 10^{-10}
\end{equation}
whereas a conservative combination of data-driven estimates yielded~\cite{Aoyama:2020ynm}
\begin{equation}
    \damuhad = 693.1 \, (4.0) \times 10^{-10}.
\end{equation}
The lattice result was at least partly corroborated by other recent lattice computations~\cite{Ce:2022kxy, ExtendedTwistedMass:2022jpw, Blum:2023qou} and, moreover, data-driven estimates using hadronic $\tau$-decays are close to lattice results \cite{Masjuan:2023qsp}. The most recent measurement of the $2\pi$ final state at CDM\nobreakdash-3~\cite{CMD-3:2023alj} compounded the mystery, as it conflicts with older measurements including CDM\nobreakdash-2~\cite{CMD-2:2006gxt}.

With these issues in mind, we wish to reconsider the statistical methodology for inferring the $R$-ratio from noisy data. As we shall discuss in \cref{sec:existing_data_driven}, the existing approaches use carefully constructed but ad hoc techniques and closed-source software, and consider uncertainties in a frequentist framework. The data-driven approach, though, is connected to  common problems in data-science and statistics: modeling an unknown function (here the $R$-ratio) and managing the risks of under- and over-fitting. In \cref{sec:tgp}, we describe how we tackle these issues using Gaussian processes --- flexible non-parametric statistical models --- and marginalization of the model's hyperparameters.\footnote{Gaussian processes were previously used to smear measurements of the $R$-ratio~\cite{Hansen:2019idp,ExtendedTwistedMassCollaborationETMC:2022sta} to facilitate a comparison with lattice predictions, as energy-smeared predictions are obtainable from lattice QCD. They were not, however, used to model the $R$-ratio itself.} This allows coherent uncertainty quantification and regularizes the wiggliness of the $R$-ratio, which helps prevent the model from over-fitting the noisy data. 
Our algorithm is implemented in our public \href{https://github.com/andrewfowlie/kingpin}{\code{kingpin}} package documented in a separate paper~\cite{kingpin}. We focus on modeling choices and developing a tool for principled modeling of the $R$-ratio; our estimates are supplementary to existing ones and we don't attempt to match previous comprehensive estimates in all respects. We don't anticipate dramatic differences with respect to previous findings; however, careful modeling of the $R$-ratio is important because $\mathcal{O}(1\%)$ changes in the HVP contribution or finding that the uncertainty was underestimated could resolve tension with the experimental measurements and lattice predictions. 
We present predictions from our model for $\damuhad$ and $\dalphahad$ in \cref{sec:results}. Finally, we conclude in \cref{sec:conclusions}. 

\section{Existing data-driven methods}\label{sec:existing_data_driven}

We now briefly review two data-driven methods for calculating $\damuhad$. First, the DHMZ approach~\cite{Davier:2010rnx,Davier:2017zfy,Davier:2019can}, which employs \code{HVPTools}, a private software package that combines and integrates cross-section data from $e^{+}e^{-}\to\text{hadrons}$. For each experiment, second-order polynomial interpolation is used between adjacent measurements to discretize the results into small bins (of around $1\,\text{MeV}$) for later averaging and numerical integration. The HVP contributions are estimated in a frequentist framework. To ensure that uncertainties are propagated consistently, pseudo-experiments are generated and closure tests with known distributions are performed to validate the data combination and integration. If the results from different experiments are locally inconsistent, the uncertainty of the combination is readjusted according to the local $\chi^{2}$ value following the well-known PDG approach~\cite{Workman:2022ynf}.

The second method is the KNT approach~\cite{Aoyama:2020ynm,Keshavarzi:2018mgv,Keshavarzi:2019abf}, which performs a data-driven compilation of hadronic $R$-ratio data to calculate the HVP contribution. It first selects the data to be used and then bins the data using a clustering procedure to avoid over-fitting. The clustering procedure determines the optimal binning of data for all channels into a set of clusters based on the available local data density. The optimal clustering criteria are shown in \refcite{Aoyama:2020ynm}. As of \refcite{Keshavarzi:2018mgv}, the KNT compilation uses an iterated $\chi^{2}$ fit to achieve the actual combination. This new method ensures that the covariance matrix is re-initialized at each iteration. The motivation of this procedure is to avoid bias. The fit results in the mean $R$-ratio for each cluster and a full covariance matrix containing all correlated and uncorrelated uncertainties. Combined with trapezoidal integration, these are used to determine channel-by-channel contributions to $\damuhad$.


The DHMZ and KNT approaches are both data-driven methods that estimate $\damuhad$ in a frequentist framework, using privately curated databases of measurements, and in-house custom codes and techniques to avoid over-fitting. The differences between the two methods are not only evident in their distinct compilation targets --- the DHMZ approach combines and integrates cross-section data from $e^{+}e^{-}\to\text{hadrons}$, while the KNT approach performs a data-driven compilation of the hadronic $R$-ratio. Furthermore, discernible disparities emerge in their respective data handling procedures, encompassing data selection, data combination, and the propagation of uncertainties. Each method has its own strengths and limitations. While DHMZ and KNT approaches do not exhibit significant differences in estimating the central value of \damuhad, there are significant disparities in the resulting uncertainties and the shapes of the combined spectra.

\section{Treed Gaussian process}\label{sec:tgp}

\subsection{Gaussian processes}

In our data-driven approach, we model the unknown $R$-ratio with Gaussian processes (GPs; \cite{williams-rasmussen-book, mackay2003information}). A GP generalizes the Gaussian distribution. Roughly speaking, whereas a Gaussian describes the distribution of a scalar and a multivariate Gaussian describes the distribution of a vector, a GP describes the distribution of a function --- an infinite collection of variables $f(x)$ indexed by a location $x$. Any subset of the random variables are correlated through a multivariate Gaussian. The degree of correlation between $f(x)$ and $f(x^\prime)$ governs the smoothness of $f(x)$ and is set by a choice of kernel function, $k(x, x^\prime)$.  

Just as a GP generalizes a Gaussian distribution of scalars or vectors to a distribution of functions, it allows us to generalize inference over unknown scalars or vectors to inference over unknown functions. Suppose we wish to learn an unknown function. Because a GP describes the distribution of a function, it can be used as the prior for the unknown function in a Bayesian setting. This prior distribution can be updated through Bayes' rule by any noisy measurements or exact calculations of the values of $f$ at particular locations $x$. In this paper we will update a GP for the $R$-ratio by the noisy measurements of the $R$-ratio. We use \code{celerite2}~\cite{celerite} for ordinary GP computations.

The kernel function is usually stationary, that is, depends only on the Euclidean distance between locations,
\begin{equation}
    k(x, x^\prime) = k(|x - x^\prime|).
\end{equation}
Once a particular form of stationary kernel has been chosen, a GP can be controlled by three hyperparameters: a constant mean $\mu$, 
\begin{equation}
    \mean{f(x)} = \mu,
\end{equation}
and a scale $\sigma$ and length $\ell$ that govern the covariance,
\begin{equation}
    \cov{f(x), f(x^\prime)} = \sigma^2 k\left(\frac{|x - x^\prime|}{\ell}\right).
\end{equation}
The scale controls the size of wiggles in the function predicted by the GP. The length determines the length scale over which correlation decays and hence the number of wiggles in an interval. For Gaussian kernels, by Rice's formula~\cite{rice} the expected number of wiggles per unit distance scales as $1 / \ell$. These three hyperparameters can substantially affect how well a GP models an unknown function. In a fully Bayesian framework, the hyperparameters are marginalized. This automatically weights choices of hyperparameter by how well they model the data and alleviates overfitting. The wigglines is regularized and the fit needn't pass through every data point.

\subsection{Treed Gaussian processes}

The GPs described thus far are stationary --- they model all regions of input space identically. To allow for non-stationary structure in the $R$-ratio, we use a treed-GP~(TGP; \cite{tgp,gramacy2020surrogates}). This is necessary as we know that the $R$-ratio contains narrow features such as resonances. In a TGP, the input space is partitioned using a binary tree. The predictions in each partition are governed by a different GP with independent hyperparameters. The number and locations of partitions are modeled using the so-called CGM prior~\cite{chipman-cart}.

\begin{figure}[t!]
    \centering
    \includegraphics[width=0.98\textwidth]{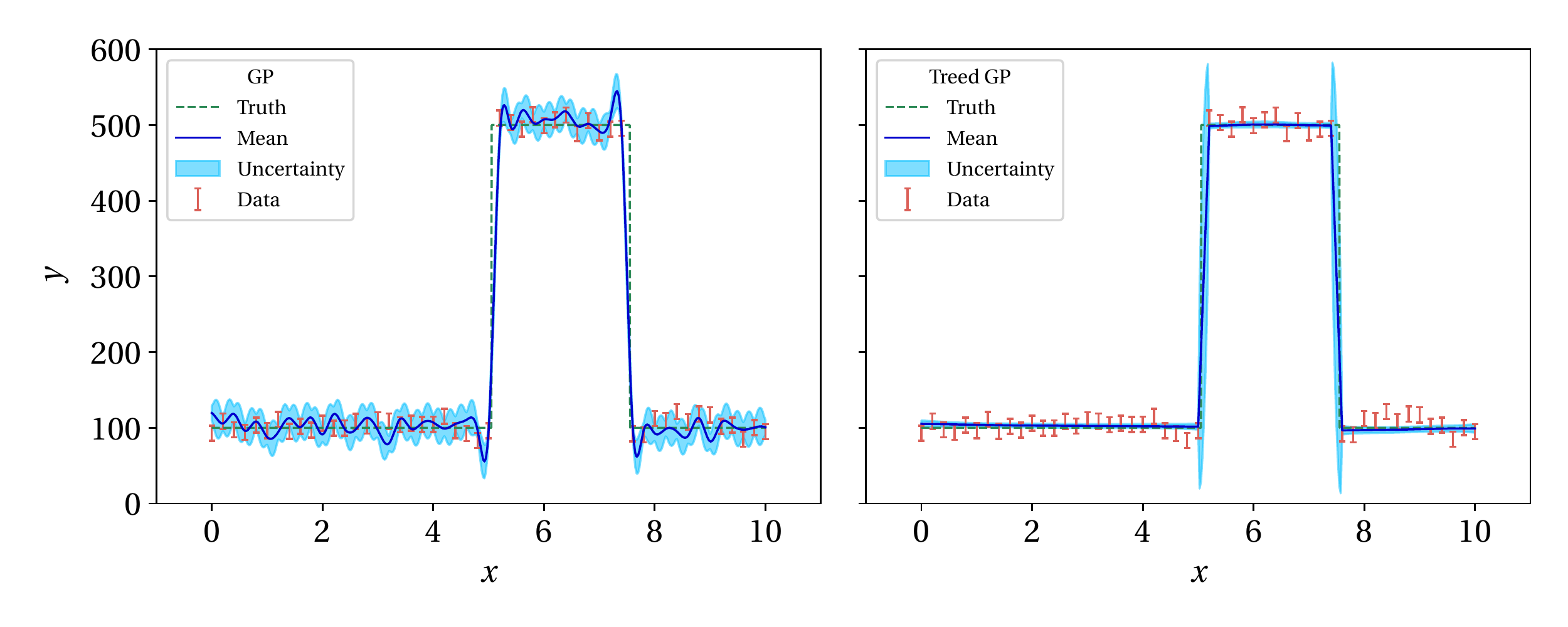}
    \caption{The GP (left) predicts a wiggly fit to the straight line sections due to the step. The TGP (right) automatically addresses the issue by partitioning the input space.}
    \label{fig:step_function}
\end{figure}

The difference in predictions between a GP and our TGP is illustrated in \cref{fig:step_function}. In this illustration we consider evenly spaced noisy measurements of a function that contains a step. The GP (left) models the data poorly, as to accommodate the sudden step, the covariance between input locations must be weak which results a wiggly fit to the straight-line sections. The TGP (right) automatically partitions the input space allowing it to model the distinct straight-line sections separately. At the jumps before and after the step, the TGP predicts the function with substantial uncertainty. This is satisfying since the data points change dramatically across those regions  of input space and do not indicate what might happen inside them.

TGPs build on ideas such as CART~\cite{chipman-cart}, treed models generally~\cite{chipman-treed} and partitioning~\cite{denison-partitioning}, and are similar to piece-wise GPs~\cite{kim-piecewise} and a recent proposal in machine learning~\cite{lederer2020real}. Alternative approaches to non-stationarity include non-stationary kernel functions~\cite{10.5555/2981345.2981380}, Deep GPs~\cite{pmlr-v31-damianou13a,2015arXiv151102222W} where non-stationarity is modeled through warping, and hierarchical models of GPs~\cite{6958906,pmlr-v51-heinonen16}. There is valuable discussion and comparison of these approaches in \refcite{sauer2022active,2022arXiv220402904S} and this remains an active area. As well as addressing non-stationarity in unknown functions, these approaches address heteroscedastic noise in our measurements.

Our approach is fully Bayesian --- we marginalize the GP hyperparameters and tree structure. This decreases the risk of over- or under-fitting the noisy data and smooths the partitions between GPs. We perform marginalization numerically using reversible jump Markov Chain Monte Carlo (RJ-MCMC;~\cite{green-rjmcmc}. For reviews see \refcite{green-hastie-review, hastie-green-review, sisson-review}). This is a generalization of MCMC that works on parameter spaces that don't have a fixed dimension --- this is vital because the number of GPs and thus the total number of hyperparameters isn't fixed. Navigating the tree structure requires special RJ-MCMC proposals --- such as growing, pruning and rotating the tree --- that are described in \refcite{tgp}.

\subsection{Integration}

The idea of modeling integrals through GPs was originally known by Bayes-Hermite quadrature~\cite{OHAGAN1991245}, and later discussed under the names of Bayesian Monte Carlo~(BMC; \cite{rasmussen2003bayesian}) and Bayesian quadrature or cubature~\cite{pmlr-v108-fisher20a,Jagadeeswaran2019}; see \refcite{briol2015probabilistic} for a review. Suppose we wish to compute an integral of the form,
\begin{equation}\label{eq:integral}
I = \int C(x) f(x) \, \text{d} x 
\end{equation}
where $f(x)$ is the estimated function and $C(x)$ is a known function. BMC provides an epistemic meaning to errors in quadrature estimates of theses integrals, such as
\begin{equation}\label{eq:approx}
I \approx \sum_i C(x) f(x_i) \Delta x_i,
\end{equation}
because we may make inferences on $I$ through our statistical model for $f(x)$. In cases in which the function $C(x)$ and choice of kernel lead to intractable computations, there is an additional discretization error in BMC inferences as the GP predictions are evaluated on a finitely-spaced grid. This is known as approximate Bayesian cubature~\cite{briol2015probabilistic}. This additional error may be neglected when the integrand is approximately linear between prediction points. We will use a TGP to model an integrand. Although trees have been proposed in BMC~\cite{NEURIPS2020_3fe94a00}, they haven't previously been directly combined with GPs in this way.

\subsection{Sequential design}

After completing inference of an unknown function with the data at hand, one may wish to know what data to collect next. This problem is known as sequential design or active learning. Broadly speaking, this is a challenging question and greedy approaches that make optimal choices one step at a time are easier to implement. We thus consider a variant of Active Learning Mackay (ALM; \cite{mackay-alm,seo-alm}).


Following the approach in \refcite{WEI2020113035}, we consider the location that contributes most to the uncertainty in the $\damuhad$ and $\dalphahad$ integrals to be an optimal location at which to perform more measurements. For an integral of the form \cref{eq:integral}, we compute
\begin{equation}\label{eq:alm}
x_\text{ALM} = \argmax \nolimits_x \left[ \int C(x) \, \cov{f(x), f(y)} \, C(y) \, \text{d} y \right].
\end{equation}
See \refcite{2008arXiv0805.4359G,gramacy2020surrogates} for further discussion.

\begin{figure}[th!]
\centering
\includegraphics[width=0.99\textwidth]{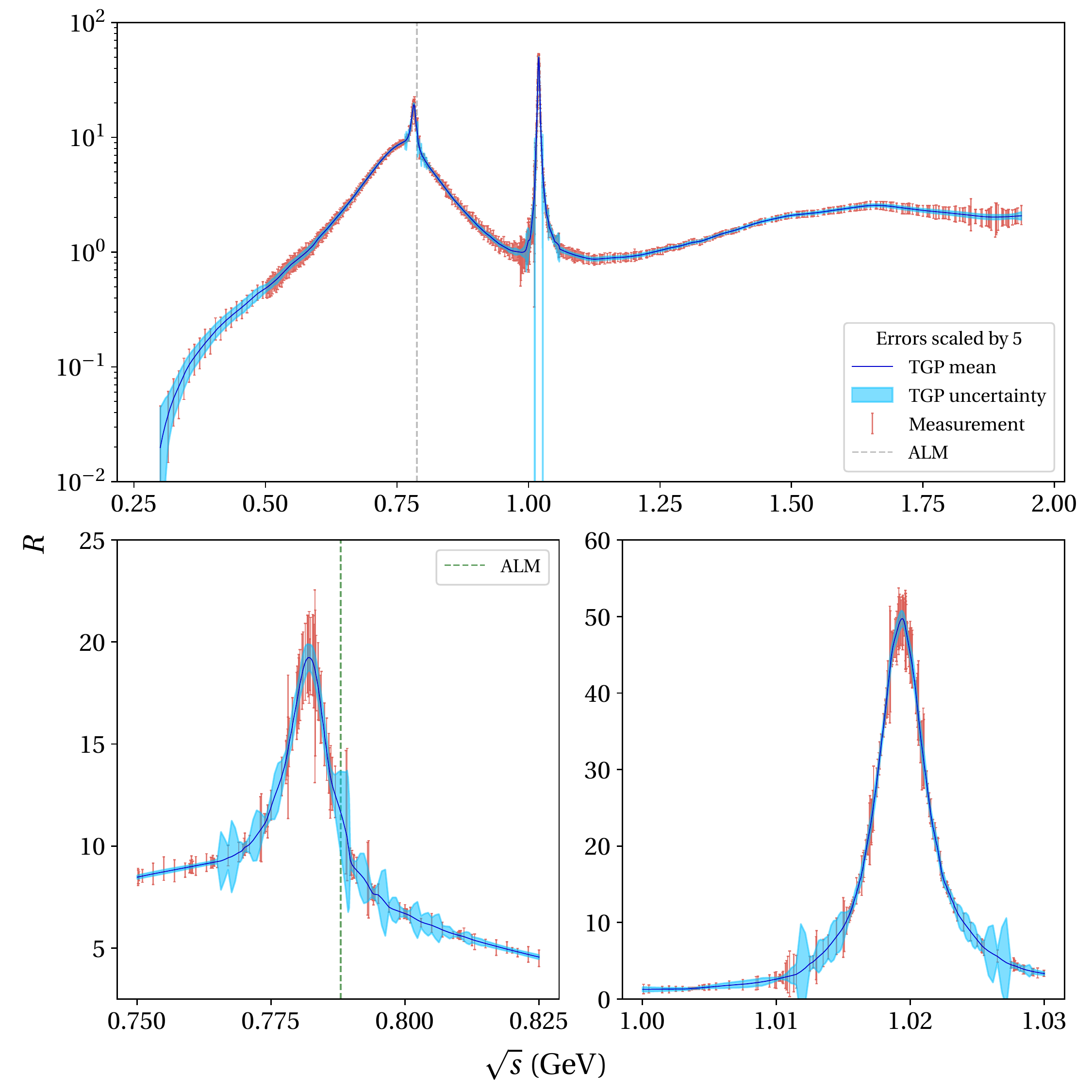}
\caption{Predicted $R$-ratio from the TGP model. The experimental errors and uncertainty in the TGP predictions are scaled by $5$ and the $\rho$ -- $\omega$ and $\phi$ resonances are plotted separately for visibility.}
\label{fig:tgp_result}
\end{figure}

\section{Results}\label{sec:results}

\subsection{Data selection}

We investigated the public dataset from the Particle Data Group (PDG;~\cite{Workman:2022ynf,pdg_url}. See also \refcite{Ezhela:2003pp} for further details), which primarily comprises data on the inclusive $R$-ratio, asymmetric statistical errors, and point-to-point systematic errors from electron-positron annihilation to hadrons at different CM energies. Certain CM energies may have multiple point-to-point systematic uncertainties stemming from different sources. We symmetrized errors and combined systematic ($\tau$) and statistical errors ($\sigma$) in quadrature:
\begin{align}
\tau^2 = \sum_{i} \tau_i^2 &\quad\text{where}\quad
\tau_i = \frac{1}{2} (\tau_{\text{up}} + \left|\tau_{\text{down}} \right|),\\
\sigma_\text{total} = \sqrt{\sigma^2 + \tau^2}  &\quad\text{where}\quad \sigma = \frac{1}{2} (\sigma_\text{up} + \left|\sigma_\text{down}\right|).
\end{align}
We selected $859$ data points inside the CM energy interval $0.3$ -- $1.937\gev$. This interval was selected to facilitate a comparison with \refcite{Keshavarzi:2018mgv}. The maximum $\sqrt{s} = 1.937 \gev$ was chosen as it is the point at which summing exclusive $R$-ratio data becomes unfeasible and perturbative QCD may be reliable. The minimum $\sqrt{s}=0.3 \gev$ was the minimum energy in the public PDG dataset.

To model this data set, we utilized a treed Gaussian process (TGP), as described in \cref{sec:tgp}. Besides selecting the data, we must specify the locations at which we want to predict the $R$-ratio. In our study, we predicted at every input location and at two uniformly spaced locations between every pair of consecutive input locations.

\subsection{Computational methods and modelling choices}

We model the $R$-ratio by a TGP in which the input space is divided into partitions using a binary tree. Each partition in our TGP is governed by a mean, $\mu$, and a Matérn-3/2 kernel with independent scale, $\sigma$, and length, $\ell$, hyperparameters. We use a uniform prior between $0$ and $150$ for the mean, a uniform prior between $0$ and $500$ for the scale, and a uniform prior between $0\gev$ to $5\gev$ for the length. These choices were motivated by the maximum measured $R$-ratio and the CM interval $0.3$ -- $1.937\gev$ under consideration. Following \refcite{tgp}, the structure of the tree itself is controlled by a CGM prior with hyperparameters $\alpha = 0.5$ and $\beta = 2$; see \refcite{chipman-cart} for explanation of these parameters. These choices favor smaller and more balanced trees.

We marginalize the tree structure and hyperparamters using RJ-MCMC. To improve computational efficiency, we thin the chains by a factor of four and only compute predictions for the states in the thinned chains. This reduces the computational time  but only slightly reduces the effective sample size as the states in the unthinned chain are strongly correlated. We run RJ-MCMC for 300\,000 steps but discard 5000 burn-in steps to minimize bias from the beginning of the chain. For computational efficiency and following a multistart heuristic, we run 10 chains in parallel and combine them.

\subsection{Predictions}

\begin{figure}[th!]
\centering
\includegraphics[width=0.7\textwidth]{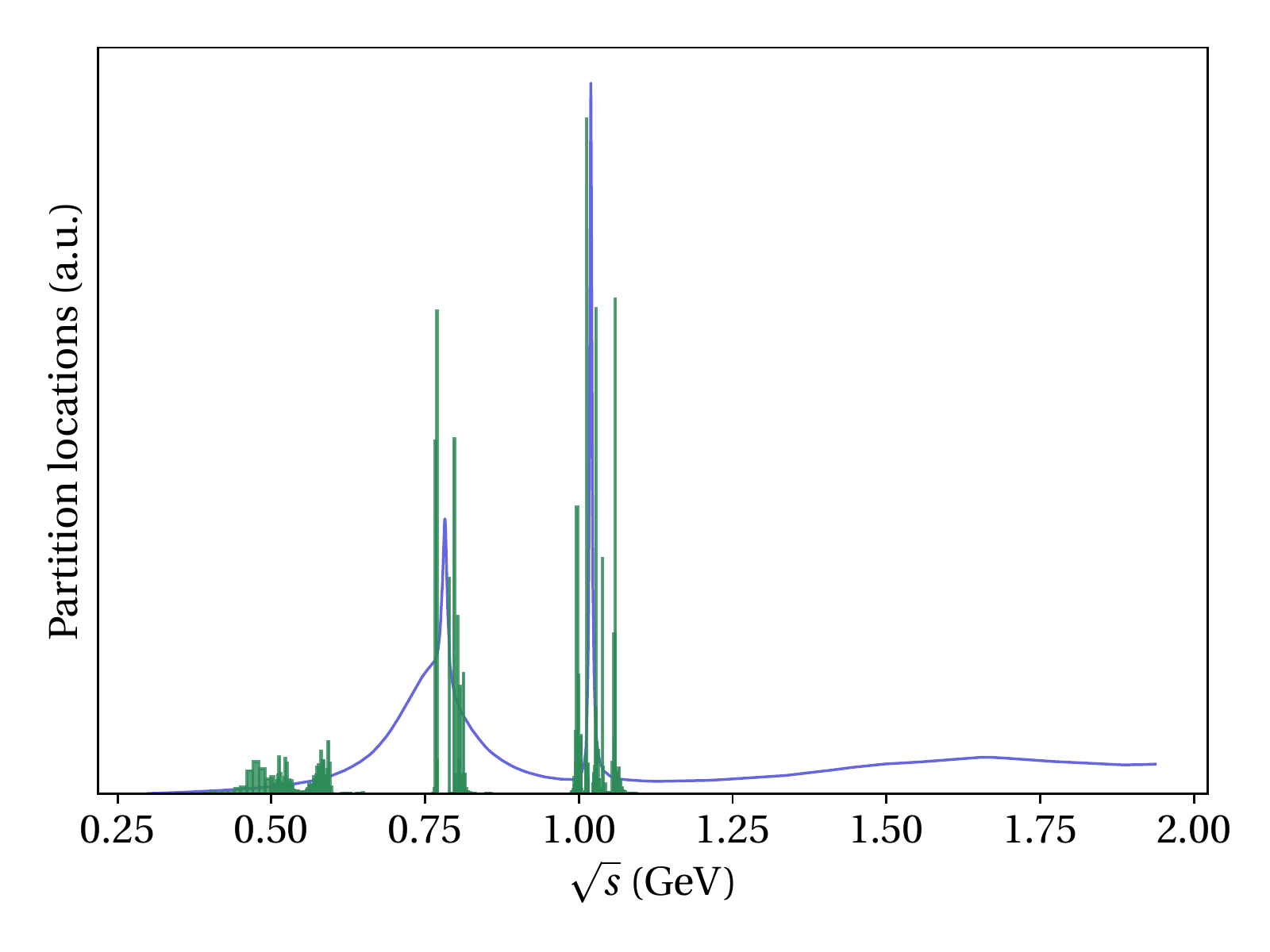}
\caption{Histogram of locations of partition edges in the TGP model. The mean prediction for the $R$-ratio is shown for reference (blue).}
\label{fig:hist}
\end{figure}

\begin{figure}[th!]
\centering
\includegraphics[width=0.99\textwidth]{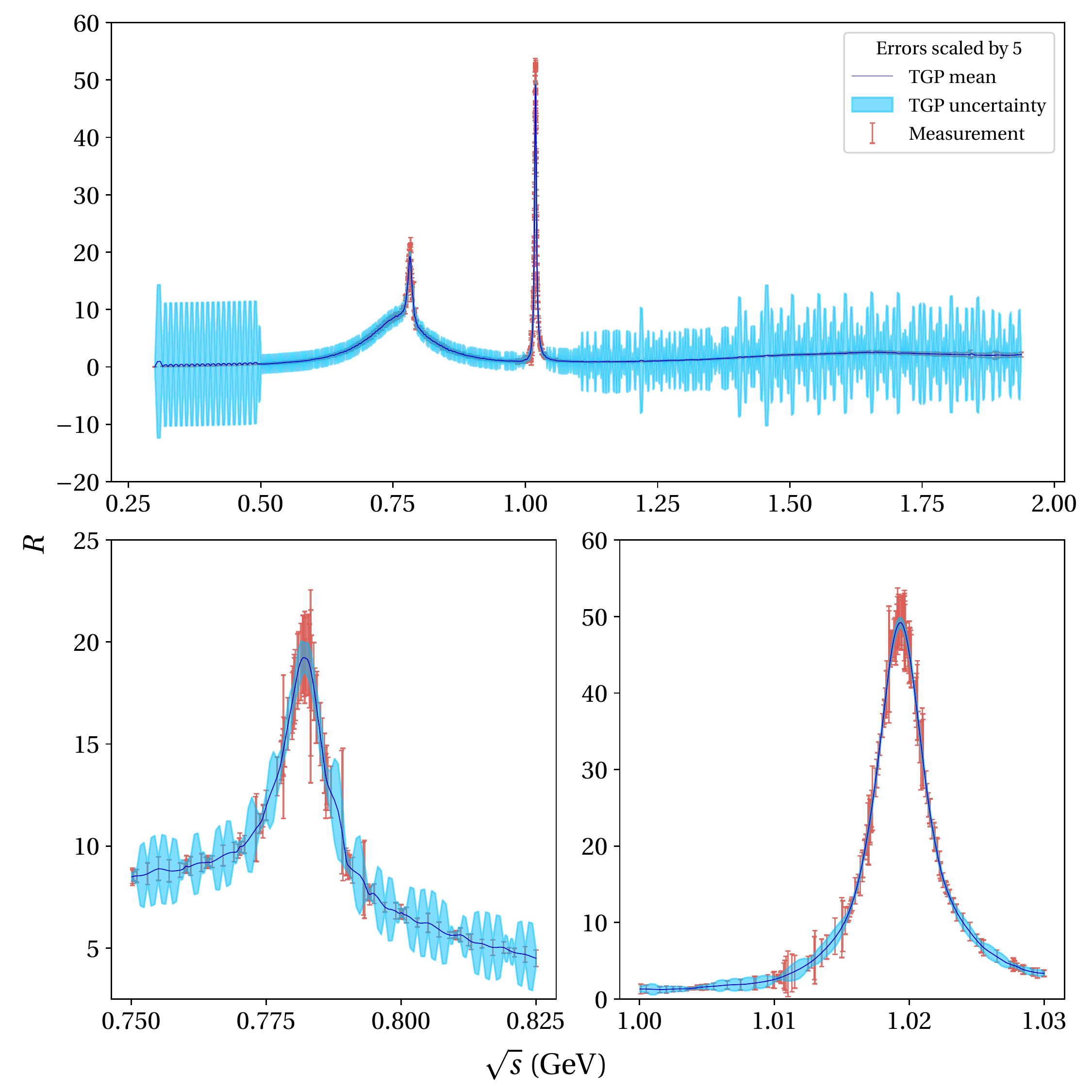}
\caption{Similar to \cref{fig:tgp_result}, though showing results from an ordinary GP.}
\label{fig:gp_result}
\end{figure}

The predictions from our TGP model for the $R$-ratio are shown in \cref{fig:tgp_result} as a mean and an error band. The mean predictions pass smoothly around the data points without any undue fluctuations near the data points that are characteristic of over-fitting. The $\rho$ -- $\omega$ and $\phi$ resonances are typically fitted by their own tree partitions with separate hyperparameters. They aren't forced to be as smooth as the rest of the spectrum and appear well-fitted. Our model predictions are noticeably more uncertain in regions with fewer or noisy measurements. We identify no anomalous features and tentatively conclude that the RJ-MCMC marginalization adequately converged. We ran standard MCMC diagnostics on the mean of the $R$-ratio using \code{ArviZ}~\cite{arviz_2019}; finding $n_\text{eff} \simeq 600$ bulk effective samples  
and Gelman-Rubin diagnostic $1.01$~\cite{2019arXiv190308008V}. There are typically five or six partitions, as the two peaks and the three flatter regions are modeled separately as shown in \cref{fig:hist}. In \cref{fig:gp_result} we show the result when using an ordinary GP. To accommodate the narrow peaks in the measured $R$-ratio, the GP model permits substantial wiggles between data points, especially where the data points are sparse.

The TGP model outputs the mean, $\mean{R_i}$, and covariance, $\cov{R_i, R_j}$, of the $R$-ratio at the prediction locations $\sqrt s_i$. The mean function represents the expected or average output value for a given input value, while the covariance function represents the covariance between predictions at different CM energies. As the RJ-MCMC can be computationally expensive and time-consuming, we saved these results to disk and made them publicly available \cite{code}. We used the mean and covariance predictions for $R$ in combination with the dispersion integrals to predict contributions to $\damuhad$ and $\dalphahad$ from the CM energy interval of $0.3$ -- $1.937\gev$. As $\damuhad$ and $\dalphahad$ are linear functions of $R$, we propagate $\mean{R_i}$ and $\cov{R_i, R_j}$ to obtain predictions. In all subsequent integration processes, we employ the trapezoidal rule~\cite{press1992numerical}, and in subsequent formulae $\sqrt{s}$ denotes the locations of our TGP predictions rather than the locations of the measurements.

The calculation of $\damuhad$ is based on \cref{eq:HVP}. However, since the independent variable in this case is the CM energy, a simple deformation of \cref{eq:HVP} is necessary,
\begin{equation}
\damuhad = \frac{2\alpha^{2}}{3\pi^{2}} \int_{m_\pi^2}^\infty \frac{K(s) R(\sqrt{s})}{\sqrt{s} } \, \text{d}\sqrt{s}
\label{HVP1}
\end{equation}
Then we calculate the value of $\damuhad$ through numerical quadrature,
\begin{equation}\label{eq:sum}
\damuhad = \sum_{i} C^\hvp_i R_i
\end{equation}
where we defined
\begin{equation}
    C^\hvp_i \equiv \frac{2\alpha^{2}}{3\pi^{2}} \frac{K(\sqrt{s}_i)}{\sqrt{s}_i} w_i
\end{equation}
where $w_i$ are the quadrature weights. We use the trapezoid rule such that
\begin{equation}
    w_i = \frac12\begin{cases}
         \sqrt{s}_{2} - \sqrt{s}_{1} & i = 1\\
         \sqrt{s}_{N} - \sqrt{s}_{N-1} & i = N\\
         \sqrt{s}_{i+1} - \sqrt{s}_{i-1} & \text{otherwise.}
        \end{cases}   
\end{equation}
From \cref{eq:sum} and by linearity, the mean can be found through
\begin{equation}\label{eq:mean}
\mean{\damuhad} = \sum_{i} C^\hvp_i \mean{R_i},
\end{equation}
and from the covariance matrix for predictions of the $R$-ratio, the uncertainty in our prediction of $\damuhad$ can be calculated using
\begin{equation}\label{eq:var}
    \var{\damuhad} = \var{\sum_{i} C^\hvp_i R_i}
    = \sum_{i, j=1}^{n} C^\hvp_i\, C^\hvp_j \,\cov{R_i, R_j}.
\end{equation}
We compute $\dalphahad$ similarly using,
\begin{equation}
\dalphahad = \frac{2\alpha M_Z^{2}}{3\pi}\fint\limits_{m_{\pi }^{2}}^{\infty}\frac{R(\sqrt{s})}{\sqrt{s}(M_Z^{2}-(\sqrt{s})^{2})}\text{d}\sqrt{s}.
\label{alpha1}
\end{equation}
We use \cref{eq:mean,eq:var} though with coefficients,
\begin{equation}
C^\text{had}_i = \frac{2\alpha M_Z^{2}}{3\pi} \frac{w_i}{\sqrt{s}_i\left[M_Z^{2}- s_i\right]}.
\end{equation}
Because our calculation is performed at CM energies from $0.3$ -- $1.937\gev$, the principal-value prescription does not need to be considered.

For the sake of comparison and to verify parts of our tool-chain, we calculate $\damuhad$ and $\dalphahad$ naively without utilizing a TGP. We consider a naive model that at the locations of the measurements of the $R$-ratio predicts
\begin{equation}
    R_i = \hat R_i \pm \sigma_i
\end{equation}
where $\hat R_i$ are the central values and $\sigma_i$ are the errors of the measurements. In this naive model there is no covariance between predictions, that is, $\cov{R_i, R_j} = 0$ for $i \neq j$. \Cref{eq:mean,eq:var} apply to this simple case, although it should be noted that in this case $\sqrt{s}$ are a series of data points, whereas in the TGP $\sqrt{s}$ are the chosen prediction locations. 

The results of the above calculations are summarized in \cref{results}. We show the predictions from  \benchmarki and \benchmarkii  for comparison, which are found by summing data-based exclusive channels in tables 2 and 1, respectively, and combing errors in quadrature.\footnote{Compared to \benchmarki, several channels were updated and two new channels measured by CMD-3~\cite{CMD-3:2019ufp} were included in \benchmarkii.}
We see that the TGP prediction for $\damuhad$ is smaller than predictions from the naive model, \benchmarki and \benchmarkii. This would make tension between data-driven estimates and lattice QCD and the experimental measurements worse. The uncertainties in our TGP predictions are nearly identical to those from the naive model --- we explain this similarity in uncertainties in \cref{sec:sim} --- though substantially smaller than those from \benchmarki and \benchmarkii. We don't anticipate that the smaller TGP uncertainties are a consequence of the TGP model itself; rather, \benchmarki and \benchmarkii are based on a different dataset and treatment of systematics. For example, they include uncertainties from vacuum polarization (VP) effects and final-state radiation (FSR) that we omit. We thus find no clear evidence of mismodelling or that our more careful modeling can shed light on the tension between data-driven estimates, lattice estimates and experiments. It is possible, however, that for an identical dataset to that in \benchmarki and \benchmarkii, the TGP predictions could be greater than \benchmarki and \benchmarkii --- the impact of reducing overfitting with a TGP could work in the opposite direction in that dataset.

\begin{table}[t!]
    \centering
    \begin{tabular}{rrrrr}
    \toprule
         & Treed GP Model & Naive Model & \benchmarki & \benchmarkii\\
    \midrule
         $\damuhad \times 10^{10}$  & $636.34 \pm 0.83$ & $637.73 \pm 0.84$ & $638.59 \pm 2.35$ & $638.25 \pm 2.32$ \\    
         $\dalphahad\times10^{4\phantom{0}}$ & $57.28 \pm 0.05$ & $57.41 \pm 0.05$ & $57.83 \pm 0.29$ & $ 57.88\pm 0.29$ \\
    \bottomrule
    \end{tabular}
    \caption{Contributions to $\damuhad$ and $\dalphahad$ in the CM energy range of $0.3$ -- $1.937\gev$ from our TGP model, a naive model, \benchmarki and \benchmarkii.}
    \label{results}
\end{table}

\subsection{Sequential design}

We may use our TGP result to identify locations that contribute most to the uncertainty in the $\damuhad$ and $\dalphahad$ predictions and where future measurements would be most beneficial. For both $\damuhad$ and $\dalphahad$, the ALM estimate from \cref{eq:alm} yields
\begin{equation}
    {\sqrt s}_\text{ALM} = 0.788 \gev.
\end{equation}
This lies near noisy measurements after the $\rho$ -- $\omega$ resonance; see \cref{fig:tgp_result}. Besides lying close to noisy measurements, the uncertainty at this location is substantial because it is a boundary between partitions of the TGP --- the behavior of the function changes abruptly here and so is hard to predict.

\subsection{Correlation}

Lastly, let us consider the relationship between the predictions for $\damuhad$ and $\dalphahad$. From \cref{eq:HVP,eq:alpha_had}, we observe that the dispersion integral formulas used to calculate $\damuhad$ and $\dalphahad$ both involve the $R$-ratio. To quantify this relationship, covariance can be used to measure the correlation between two variables. The sign of covariance indicates whether the trends between the two variables are consistent. The correlation coefficient is usually utilized to reflect the strength of the correlation between two variables. Thus, to gain more understanding about the relationship between $\damuhad$ and $\dalphahad$, we computed their covariance and correlation,
\begin{align}
\cov{\damuhad,\dalphahad} &= \sum_{i,j=1}^{n} C^\hvp_i \, C^\text{had}_j \, \cov{R_i, R_j} 
\label{Cov}\\
\corr{\damuhad,\dalphahad} &= \frac{\cov{\damuhad,\dalphahad}}{\sqrt{\var{\damuhad} \var{\dalphahad}}}
\label{Cof}
\end{align}
where the (co)variances were computed under the TGP as described. As anticipated, we obtained a positive correlation between the two. Specifically, when $\dalphahad$ increases, the value of $\damuhad$ also increases, and vice versa. The calculated correlation coefficient was $\rho \simeq 0.8$, which is close to $1$, quantifying the strong correlation between $\damuhad$ and $\dalphahad$.

\section{Discussion and conclusions}\label{sec:conclusions}

The BNL and FNAL measurements of the anomalous magnetic moment of the muon disagree with the Standard Model (SM) prediction by more than $4\sigma$. This has led to renewed scrutiny of new physics explanations and the SM prediction. With that as motivation, we extracted the hadronic vacuum polarization (HVP) contributions, $\damuhad$, from electron cross-section data using a treed Gaussian process (TGP) to model the unknown $R$-ratio as a function of CM energy. This is a principled and general method from data-science, that allows complete uncertainty quantification and automatically balances over- and under-fitting to noisy data.

The challenges in the data-driven approach are common in data-science. A competitive estimate of $\damuhad$, however, requires domain-specific expertise, careful curation of measurements, and careful consideration of systematic errors and their correlation. This should be developed over time in collaboration with domain experts. Thus our work should be seen as preliminary and serves to explore an alternative statistical methodology based on more general principles and develop an associated toolchain. We used a dataset available from the PDG, though as noted as early as 2003 in \refcite{Ezhela:2003pp}, a more complete, documented and standardized database of measurements would allow further scrutiny of data-driven estimates of HVP.

Our analysis used about $n \approx 1000$ data points. The linear algebra operations in GP computations scale as $\mathcal{O}(n^3)$. There are computational approaches and approximations to overcome this scaling~(see e.g.,~\refcite{seeger2003fast,quinonero2005unifying,snelson2007local,8951257,hensman2013gaussian}); nevertheless, working with more complete datasets could be challenging. On the other hand, splitting data channel by channel could help the situation. For a competitive estimate, we would require careful treatment of correlated systematic uncertainties. The approach started here --- carefully building an appropriate statistical model --- naturally allows us to model systematic uncertainties. For example, through nuisance parameters for scale uncertainties or sophisticated noise models for correlated noise~(see e.g., \refcite{Delisle_2020}).  The statistical model could include, for example, a hierarchical model of systematic uncertainties accounting for ``errors on errors.''

The prediction for $\damuhad$ from our TGP model is slightly smaller than existing data-driven estimates. Thus, more principled modeling of the $R$-ratio in fact increases tension between the SM prediction and measurements for $g-2$. On the other hand, because the kernel functions were slowly-varying, the TGP model predicted $\damuhad$ with a similar uncertainty to that obtained in naive approach. This can be understood from the trade-off between variance and covariance in predictions of the $R$-ratio at different CM energies. Looking forward, by the ALM criteria, the best CM energy for future measurements was $\sqrt s \simeq 0.788\gev$ for both $\damuhad$ and $\dalphahad$, as it lies close to particularly noisy measurements of the $R$-ratio. In conclusion, we developed a statistical model for the $R$-ratio, based on general principles and publicly available toolchains. We found no indication that mismodeling the $R$-ratio could be responsible for tension with measurements or lattice predictions. We hope, however, that this work serves as a starting point for further scrutiny, principled modeling and development of associated public tools.

\section*{Acknowledgments}

AF was supported by RDF-22-02-079. We thank Peter Athron for comments and feedback.

\section*{Data and code availability}

The TGP algorithm is implemented in our public \code{Python} package \code{kingpin}~\cite{kingpin}. The dataset and codes, which use the \code{kingpin} package, for this paper are available online; see \refcite{code}.   

\appendix

\section{Similarity between TGP and naive model uncertainties}\label{sec:sim}

Here we explain why the uncertainties in the predictions for $\damuhad$ and $\dalphahad$ calculated under our TGP and in the naive method are approximately equal, 
\begin{align}\label{eq:tgp_uncertainty}
    \sigma^2_\text{TGP} &= \sum_{i,j=1}^{n} C_i C_j \cov{R_i, R_j} \\
    & \simeq \sigma^2_\text{naive} = \sum_{i=1}^n C_i^2 \sigma_i^2,\label{eq:naive_uncertainty}
\end{align}
despite quite different predictions for the $R$-ratio. We anticipated a reduction in error in our TGP as we applied prior information about correlations between prediction points. There are two effects of introducing correlation: first, as anticipated a reduction in the variance at any prediction point, that is, $\var{R_i} \ll \sigma^2_i$. Second, an increase in the correlation between prediction points, $\cov{R_i, R_j} > 0$. The uncertainty in our TGP predictions includes terms containing variance and covariance,
\begin{equation}
    \sigma^2_\text{TGP} = \sum_{i,j=1}^{n} C_i C_j \,\cov{R_i, R_j} = \sum_{i=1}^n C_i^2 \, \var{R_i} + \sum_{i \neq j} C_i C_j \,\cov{R_i, R_j}
\end{equation}
In practice we find that the decrease in the variance is almost exactly canceled by the increase in the covariance.

\begin{figure}[t!]
    \centering
    \includegraphics[width=0.99\textwidth]{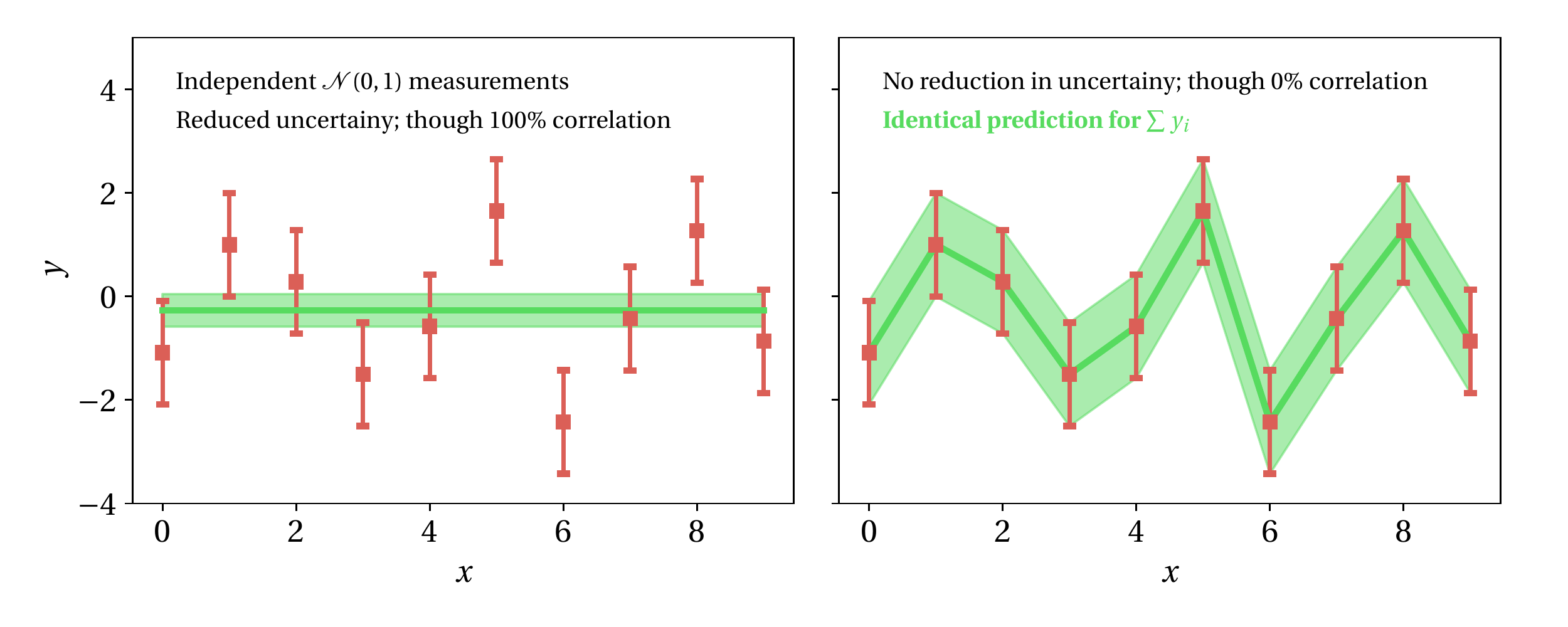}
    \caption{A horizontal line fitted to noisy data (left) and a naive model that passes through every data point (right). Despite making quite different predictions for $y$, they make identical predictions for $\sum y_i$.}
    \label{fig:sketch}
\end{figure}

To understand this effect in the simplest way, consider fitting a horizontal line ($m=0$) with an unknown intercept ($c$) through data points with errors $\sigma$ as shown in \cref{fig:sketch}. The uncertainty at any prediction point equals the uncertainty on the intercept,
\begin{equation}
    \var{y_i} = \frac{\sigma^2}{n}
\end{equation}
This is reduced from $\sigma^2$ by the $100\%$ correlation between the predictions at the $n$ data points. Now consider the uncertainty on the sum,
\begin{align}
    \var{\sum_{i=1}^n y_i} &= \sum_{i,j=1}^{n} \cov{y_i, y_j} 
            = \sum_{i=1}^n \var{y_i} + \sum_{i \neq j} \cov{y_i, y_j} \\
            &= n \frac{\sigma^2}{n} + (n^2 - n) \frac{\sigma^2}{n} = n \sigma^2
\end{align}
This is identical to the uncertainty from our naive model with no correlations that fits $y = \hat y \pm \sigma$ because the increased covariance cancels the decreased variance. Although we reduced the uncertainty at any prediction point, that reduction was offset by the covariance between predictions.

Let us create an example closer to our TGP model and demonstrate an identical effect. In a GP with fixed hyperparameters for measurements with uniform noise $\sigma$, the covariance between prediction points $X^\star$ can be expressed as
\begin{equation}
\cov{y(X^\star), y(X^\star)} = K(X^\star, X^\star) - K(X^\star, X)^{T}  \left[K(X, X) + \sigma^2\right]^{-1} K(X^\star,X)
\label{error}
\end{equation}
where $X$ are the training points, $K$ is the choice of kernel function, and $\sigma^2$ is the noise. This expression is somewhat intractable due to the inverse matrix. Thus we consider $X = X^\star$ and a simplified covariance,
\begin{equation}\label{eq:simple_kernel}
K_{ij} = a + b \delta_{ij}.
\end{equation}
With this choice, one can compute $\cov{y(X^\star), y(X^\star)}$ and the sum over its elements analytically. The covariance may be written as
\begin{equation}
\cov{y(X^\star), y(X^\star)} = A 
\begin{pmatrix}
  D           & a\sigma^{2} & \cdots & a\sigma^{2}\\
  a\sigma^{2} & D           & \cdots & a\sigma^{2}\\
  \vdots      & \vdots      & \ddots & \vdots\\
  a\sigma^{2} & a\sigma^{2} & \cdots & D\\
\end{pmatrix}
\end{equation}
where 
\begin{align}
A &\equiv \frac{\sigma^{2}}{(b+\sigma^{2})(na+b+\sigma^{2})}\\
D &\equiv b(na+b)+(a+b)\sigma^{2}.
\end{align}
Summing the elements results in,
\begin{equation}
    \sum\cov{y(X^\star), y(X^\star)} = \frac{n(na+b)\sigma^{2}}{na+b+\sigma^{2}} 
\end{equation}
where $n$ represents the size of $X^\star$. By performing a Taylor expansion about $\sigma =0$, we discover
\begin{equation}
    \sum\cov{y(X^\star), y(X^\star)} = n\sigma^{2} + \mathcal{O}(\sigma^4).
\end{equation}
Although the structure \cref{eq:simple_kernel} isn't particularly realistic, our result holds for any $a$,  including $0\%$ and $100\%$ correlation. When $C_i \approx C \equiv \text{const.}$, the uncertainties from the TGP and naive method are both around $C \sqrt{n} \sigma$. The decrease in variance and increase in covariance cancel in the TGP uncertainties. For the case in which the prediction locations are denser than the input locations, this result may hold if the TGP predictions don't change substantially between input locations.

\bibliographystyle{JHEP}
\bibliography{refs}

\end{document}